# Valley and spin pump by scattering at non-magnetic disorders

Xing-Tao An[1,2], Jiang Xiao[3], M.W.-Y. Tu[1], Hongyi Yu[1], Vladimir I. Fal'ko[4], Wang Yao[1,*]

[1]Department of Physics and Center of Theoretical and Computational Physics, University of Hong Kong, Hong Kong, China

[2]School of Science, Hebei University of Science and Technology, Shijiazhuang, Hebei 050018, China

[3]Department of Physics and State Key Laboratory of Surface Physics, and Collaborative Innovation Center of Advanced Microstructures, Fudan University, Shanghai 200433, China

[4]National Graphene Institute, and School of Physics and Astronomy, University of Manchester, Manchester, UK

[*]Correspondence: wangyao@hku.hk

**In solid, the crystalline structure can endow electron an internal degree of freedom known as valley, which characterizes the degenerate energy minima in momentum space. The recent success in optical pumping of valley polarization in 2D transition metal dichalcogenides (TMDs) has greatly promoted the concept of valley-based informatics and electronics. However, between the demonstrated valley polarization of transient electron-hole pair excitations and practical valleytronic operations, there exist obvious gaps to fill, among which is the valley pump of long-lived charge carriers. Here we discover that the quested valley pump of electrons or holes can be realized simply by scattering at the ubiquitous nonmagnetic disorders, not relying on any specific material property. The mechanism is rooted in the nature of valley as a momentum space index: the intervalley backscattering in general has valley contrasted rate due to the distinct momentum transfers, causing a net transfer of population from one valley to another. As examples, we numerically demonstrate the sizable valley pump effects driven by charge current in nanoribbons of monolayer TMDs, where the spin-orbit scattering by non-magnetic disorders also realizes spin pump for the spin-valley locked holes. Our finding points to an unexpected new opportunity towards valley-spintronics, turning disorders from a deleterious factor to a resource of valley and spin polarization.**

## Introduction

The energy dispersion of electrons in crystalline solids usually has degenerate minima located at well-separated momentum space points, known as valleys. For low energy electrons, the valleys span an internal quantum degree of freedom, just like the spin. Similar to the potential use of spin polarization in spintronics, information can be represented by the valley polarization configurations, i.e. unequal population distribution among the degenerate valleys. Schemes to induce or manipulate valley polarization for potential valleytronics have been explored in various systems[1-10]. Unlike spintronic controls based on generic properties of spin, these valley control schemes rely on specific materials properties such as valley dependent anisotropic dispersions[1,2], and carefully engineered strains[3,4], edges[5] or defects[6-10].

Two-dimensional hexagonal crystals such as graphene and monolayer transition metal dichalcogenides (TMDs) have newly emerged as a promising semiconducting platform for exploring valleytronic applications[11-19]. These materials have a time reversal pair of valleys spanning a pseudospin-1/2, which can acquire spin-like properties that allows its manipulation similar to the spin controls. These include the valley Hall effect[11,13-17], the valley magnetic response[11,20-23], and the valley optical selection rules[12,13]. The latter, in particular, has enabled dynamic pumping of valley polarization of electron-hole pairs by circular polarized light in 2D TMDs, observed through the luminescence of these transient excitations[24-27]. Moreover, in the presence of strong spin-orbit splitting, time-reversal symmetry dictates the opposite sign of the splitting at the two valleys, introducing the effective coupling of spin with valley pseudospin. These valley controls, however, rely on inversion symmetry breaking in the 2D hexagonal lattices[28], which has limited the generalization to more platforms.

Here we discover a general mechanism to pump valley based on its generic nature as a momentum-space index, rather than extrinsically acquired properties dependent on host materials. Because of the distinct momentum transfers on a finite Fermi surface, intervalley backscattering by non-magnetic disorder can have valley-contrasted rate, which causes a net population transfer between the valleys. In a quasi-1D geometry, valley current can be pumped out from both sides of a disordered region when a charge current is driven through. Remarkably the valley pump efficiency (i.e. per charge transmission) is shown to increase with the density of sharp disorders. For spin-valley coupled carriers, such valley pump can also realize a spin pump. In monolayer TMDs, we numerically demonstrate the sizable valley pump of electrons by spin-independent scattering in zigzag nanoribbons, and the spin pump of holes by spin-orbit scattering in nanoribbons of various orientations. Our finding is an illuminating example on the advantage of exploiting valley in future electronics, where the pseudospin controllability can directly arise from its momentum nature, in addition to those extrinsically acquired spin-like properties.

## Results

### Valley pump by intervalley scattering

Consider a quasi-1D system where the energy dispersion has two well-separated valleys centered at finite momenta $K$ and $-K$. The disorder scattering for a carrier incident in valley $\tau$ ( $\tau = \pm K$ ) can be characterized by the *valley-conserved* transmission ($T_{\tau,\tau}$) and reflection coefficients ($R_{\tau,\tau}$), and the *valley-flip* ones ($T_{\tau,-\tau}$ and $R_{\tau,-\tau}$). If valley $\tau$ has $N_\tau$ sub-bands at the Fermi energy, $T_{\tau,\tau}$ ($T_{\tau,-\tau}$) and $R_{\tau,\tau}$ ($R_{\tau,-\tau}$) then refer to the overall valley-conserved (valley-flip) transmission and reflection summed over initial and final sub-bands of scattering.

The two valley-flip reflection coefficients $R_{K,-K}$ and $R_{-K,K}$ are different in general, as they correspond to intervalley scatterings with distinct momentum transfers, as illustrated in Fig. 1. The two valley-flip transmission coefficients are always identical, i.e. $T_{K,-K} = T_{-K,K}$, as these two scattering channels are conjugate of each other (c.f. Fig. 1). Thus, for a valley-unpolarized incident flux of $N$ electrons ($N = 2N_K = 2N_{-K}$), the population difference between the $K$ and $-K$ valleys in the outgoing flux is (counting both transmission and reflection),

$$P_V \equiv 2(R_{-K,K} - R_{K,-K}) \qquad (1)$$

The scattering by the ubiquitous non-magnetic disorder therefore provides a mechanism to pump valley.

The pumped valley polarization is carried by both the reflection and transmission flux. If the $K$ and $-K$ valleys are time reversal of each other (as in the case of graphene and TMDs), the two valley-conserved reflection coefficients $R_{K,K}$ and $R_{-K,-K}$ are also identical. For a valley-unpolarized incident flux, the reflection flux then carries a valley current of $j_V = -(R_{-K,K} - R_{K,-K})v_F$, while the valley current in the transmission flux is $j_V = (T_{K,K} - T_{-K,-K})v_F$, $v_F$ being the Fermi velocity. The valley currents on the two sides can be related through the sum rule $N_{\tau K} = T_{\tau K,\tau K} + R_{\tau K,\tau K} + T_{\tau K,-\tau K} + R_{\tau K,-\tau K}$. For the time reversal pair of valleys, this sum rule leads to:

$$T_{K,K} - T_{-K,-K} = -(R_{K,-K} - R_{-K,K}) = P_V/2. \qquad (2)$$

Namely, the valley current in the transmission flux has the same magnitude but opposite direction to that in the reflection flux (c.f. Fig. 1c), and the net outward valley flow from the disorder is $P_V v_F$.

A disordered region can thus be exploited as a valley source when a charge current is driven through by bias voltage or temperature gradient (Fig. 1c). The charge current, normalized by the thermodynamic driving force, is $j_C = T_{sum} v_F$ where $T_{sum} \equiv T_{K,K} + T_{-K,-K} + T_{K,-K} + T_{-K,K}$. The ratio between the outward valley flow from the disordered region and the charge current equals to $P_V/T_{sum}$, which characterizes the *valley pump efficiency*. It counts the valley population difference pumped per charge transmission. The sign of the valley pump changes when the direction of the charge current is flipped.

If the incident flux has a valley polarization $\eta$, i.e. $(1 + \eta)\frac{N}{2}$ electrons in valley $K$ and $(1 - \eta)\frac{N}{2}$ electrons in $-K$, the overall effect of intervalley scattering then depends on $\eta$. The outgoing flux (reflection plus transmission) has an average valley polarization of $\eta + \Delta\eta$, where

$$\Delta \eta = \frac{1}{N} P_V - \frac{1}{N} \Gamma_V \eta. \qquad (3)$$

$\Gamma_V \equiv 2(R_{-K,K} + R_{K,-K} + T_{-K,K} + T_{K,-K})$ here characterizes the valley depolarization by the intervalley scattering. The overall effect of intervalley scattering includes two counteracting terms: a valley pump term, and a depolarization term that is proportional to the incident valley polarization $\eta$. The competition between these two opposite effects determines whether the disorder scattering would increase or decrease valley polarization of incident carriers.

As a remarkable feature of this valley pump mechanism, the pump efficiency is expected to increase with the number of sharp disorders in the scattering region, since each disorder pumps valley polarization of the same sign when the quantum interference between the multiple scatters can be neglected (see Fig. 1c). The increase of $P_V/T_{sum}$ with the disorder number will eventually saturate as suggested by the valley depolarization term in Eq. (3).

**Valley pump of electrons in monolayer TMDs nanoribbons**

We demonstrate this valley pump in nanoribbons of monolayer TMDs. In these 2D semiconductors, the conduction and valence band edges at the $\pm K$ valleys are contributed predominantly by the three metal $d$-orbitals[29]: $d_{z^2}, d_{xy}, d_{x^2-y^2}$. Our calculation is thus based on the tight-binding model constructed with these three orbitals that describes well the band edge electrons and holes[30],

$$H = \sum_i \sum_\mu \varepsilon_{i\mu} c_{i\mu}^\dagger c_{i\mu} + \sum_{<i,j>} \sum_{\mu\nu} t_{i\mu,j\nu} c_{i\mu}^\dagger c_{j\nu}. \qquad (4)$$

Here $c_{i\mu}^\dagger$ creates an electron on orbital $\mu$ at metal site $i$ in an orthogonal basis, the sums $<i,j>$ run over all pairs of nearest-neighbor metal sites, and $t_{i\mu,j\nu}$ are the hopping integrals based on symmetry consideration (without the Slater-Koster two-center approximation) fitted from first principles band structures[30]. The spin-orbit coupling, being weak in the conduction bands, is neglected in discussing the valley pump of electrons in this section.

The disorder potential is introduced by a position dependent on-site energy: $\varepsilon_{i\mu} = \varepsilon_\mu + u \sum_l \exp(-\frac{|r_i - r_l|^2}{2d^2})$, where $l$ runs over the disorders randomly distributed in a region of length $L$ on a zigzag nanoribbon (c.f. Fig. 2a). All disorders are assumed with the same Gaussian profile of length scale $d = a$ ($a$ is the lattice constant) and strength $u = -0.5$eV. The valley-dependent scattering by the entire disordered region is calculated with the tight-binding Hamiltonian in Eq. (4), using a recursive Green's function technique [31]. The band parameters $\varepsilon_\mu$ and $t_{i\mu,jv}$ are taken from Ref [30]. To focus on the transport in the nanoribbon bulk rather than the edge, in the model study in this and the next sections, periodic boundary condition is used instead at the ribbon edge to get rid of the edge-states.

Fig. 2c shows the calculated valley-dependent transmission and reflection for disorder scattering in a zigzag MoS$_2$ nanoribbon of width $W = 27.7a$. The scattering region is of a length $L = 90a$, with the disorder density (numbers per unit cell) $n_i = 2\%$. As expected, the two valley-flip transmission coefficients $T_{K,-K}$ and $T_{-K,K}$ are always identical at all incident energy $E_F$, and so do the two valley-conserved reflection $R_{K,K}$ and $R_{-K,-K}$. The two valley-flip reflection coefficients $R_{K,-K}$ and $R_{-K,K}$ start to differ when $E_F$ is about $0.1$eV above the conduction band edge $E_c$, and the same amount of difference is found between $T_{K,K}$ and $T_{-K,-K}$, as dictated by Eq. (2).

The current induced valley pump efficiency is quantified in Fig. 2d-h. The valley pump $P_V = 2(R_{-K,K} - R_{K,-K})$, the overall transmission $T_{sum} = T_{K,K} + T_{-K,-K} + T_{K,-K} + T_{-K,K}$, and the valley depolarization rate $\Gamma_V$ are shown respectively in Fig. 2d, 2e and 2f at several disorder densities $n_i$. Quantum interference between the multiple scatters is preserved in our calculation, which complicates the pump behavior. This is evident from the sign difference of $P_V$ between the $n_i = 1\%$ and $n_i = 2\%$ curves at certain incident energy in Fig. 2d.

The valley pump per charge transmission $P_V/T_{sum}$ as a function of $n_i$ at a fixed Fermi energy is shown in Fig. 2g. Clearly, the pump becomes more efficient with the increase of disorder density. At $n_i = 5\%$, $P_V/T_{sum} \sim 30\%$, meaning that equivalently 3 electrons are pumped out in full valley polarization per 10 electrons transmitting through the disordered regions, which is a significant pump efficiency. At fixed disorder density, increasing the length of the scattering region can also enhance the pump efficiency, as shown in Fig. 2h by the plot of $P_V/T_{sum}$ as a function of $L$. The valley pump efficiency has the expected saturation behavior at large $L$, when the valley depolarization effect by the intervalley scattering starts to balance with the valley pump (c.f. Eq. (3)).

**Spin pump of spin-valley locked holes**

We consider now the spin-orbit splitting in the Bloch bands. At the time reversal pair of valleys, the spin splitting must have opposite sign. This valley dependent spin splitting effectively introduces a coupling between the spin and valley pseudospin, which makes spin pump possible even if the scatter is *non-magnetic*.

The idea can again be illustrated in the TMDs nanoribbons, where the valance band edge has a spin-splitting $\lambda$ of hundreds of meV[29]. So for band edge holes, the $K$ $(-K)$ valley only has spin up (down) in out-of-plane direction, i.e. the spin index is locked to the valley index (c.f. Fig. 3a). We focus on this spin-valley locked energy window of $\lambda$, where valley pump is a spin pump at the same time. The spin-valley locking also allows the pump effect to be explored in nanoribbons of various orientations, since the spin index can be unambiguously tracked even when the valleys start to overlap when projected to the propagation direction.

Our numerical calculations are performed with the spin-orbit coupling added to the three-band tight-binding model of monolayer TMDs[30],

$$H = \sum_i \sum_\mu \varepsilon_\mu \, c_{i\mu}^\dagger c_{i\mu} + \lambda \sum_i \sum_{\mu,\nu} c_{i\mu}^\dagger (\boldsymbol{L}_{\mu,\nu} \cdot \boldsymbol{S}) c_{i\nu} + \sum_{<i,j>} \sum_{\mu\nu} t_{i\mu,j\nu} \, c_{i\mu}^\dagger c_{j\nu}$$

$$-i\alpha \sum_{<j,l>_d} \sum_\mu c_{j\mu}^\dagger [\boldsymbol{S} \times \hat{\boldsymbol{r}}_{lj}]_z c_{l\mu}. \quad (5)$$

where the second term is the on-site interaction between orbital angular momentum $\boldsymbol{L}$ and spin $\boldsymbol{S}$, which accounts for the strong spin-valley coupling in the valence band[30].

$L_{\mu,\nu}$ denotes matrix element of $L$ between orbital $\mu$ and $\nu$. For the disorder, we consider here a non-magnetic one consisting of the spin-flip hopping between three nearest neighbor metal sites (Fig. 3a inset). Such hopping is described by the last term in Eq. (5), where $\hat{r}_{lj}$ is the unit directional vector pointing from site $j$ to site $l$, $< j, l >_d$ runs over the pairs of nearest-neighbor sites at each disorder, and $\alpha = 0.5$ eV.

Fig. 3b shows the calculated spin-flip and spin-conserved reflection coefficients, at a disorder density $n_i = 1\%$, in monolayer $MoS_2$ nanoribbons of zigzag, (4,1) and (2,1) orientations respectively. The intervalley reflection vanishes when the Fermi energy is between the edge of the first and the second sub-bands. This is because the intervalley reflection within the first sub-band is between a time-reversal pair of states, which vanishes for a non-magnetic disorder potential that preserves the time-reversal symmetry (c.f. Fig. 3a). When the two spin-flip reflection coefficients $R_{\uparrow,\downarrow}$ and $R_{\downarrow,\uparrow}$ become finite at higher energy, their magnitudes are different, which lead to the spin pump.

Fig. 3c shows $P_S/T_{sum}$, the spin pump per charge transmission, where $P_S \equiv 2(R_{\uparrow,\downarrow} - R_{\downarrow,\uparrow})$. Note that Eq. (1-3) can apply to the spin pump, with the valley index replaced by the spin index. In all three nanoribbons of different orientations, spin pump gets more efficient with the increase of disorder density. For this disorder potential, both the reflection and the spin pump efficiency get stronger in directions other than the zigzag.

**Discussions**

We have shown that the quested valley pump of electrons and holes can be realized with the ubiquitous non-magnetic disorders. Upon intervalley scattering by disorders, a valley pump effect on carriers coexists with the valley depolarization. The valley pump dominates when incident carriers have zero or small valley polarization, and the effect arises from the general dependence of the intervalley backscattering rate on the incident valley index. Remarkably, a practical source of valley polarization can thus be realized simply by passing charge current through a disordered region. The net effect of the disorder scattering is to turn a valley-unpolarized incident current into valley polarized transmission and reflection currents. Valley currents are thus pumped from the disordered region to the clean regions (i.e. lack of sharp disorders) on the two sides, where valley polarization can retain long with the absence of intervalley scattering.

Our finding also points to a radically new possibility for pumping spin as well. Unlike the conventional approaches, the presence of magnetization is not required here. Spin pumping is made possible by scattering at non-magnetic disorders in the presence of the spin-valley coupling. The valley and spin pump here arises from the generic nature of valley as a momentum space index, which can be exploited generally in metals and semiconductors with the valley degeneracy.

**Acknowledgments:** This work is mainly supported by the Croucher Foundation under the Croucher Innovation Award, the RGC (HKU17305914P) and UGC (AoE/P-04/08) of HKSAR, and HKU OYRA. X.A. is also supported by NSFC (No. 11104059 and 61176089) and HSFC (No. A2015208123). J.X. acknowledges support by NBRPC (No. 2014CB921600).

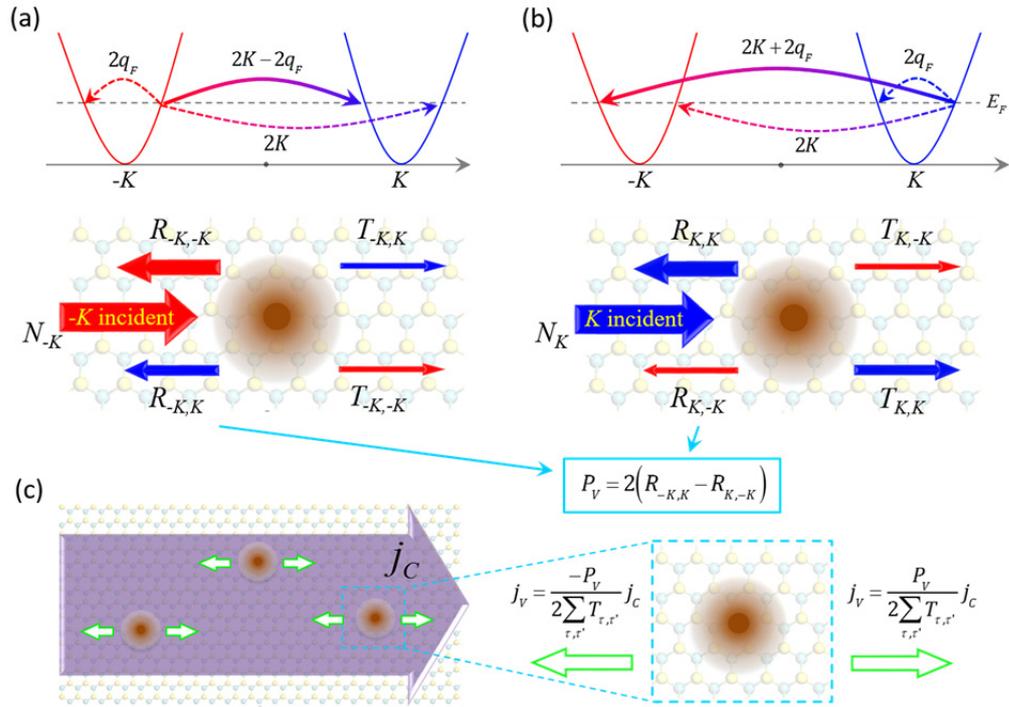

**Figure 1. Valley pump by intervalley scattering**. (a) Schematics of the momentum transfers (upper) of the scattering channels (lower) for electron incident in valley $-K$. (b) Electron incident in valley $K$. The valley-flip reflection coefficients $R_{-K,K}$ in (a) and $R_{K,-K}$ in (b) can differ because of the distinct momentum transfers (solid curved arrows in upper panels), where as the valley-flip transmission coefficients $T_{K,-K}$ and $T_{-K,K}$ always equal. For an incident flux with one electron per valley, the disorder scattering thus transfers a net population of $R_{-K,K} - R_{K,-K}$ from valley $-K$ to $K$. (c) When a charge current ($j_c$, purple arrow) is driven through, valley current ($j_V$, green arrows) is pumped out from both sides of the disorders.

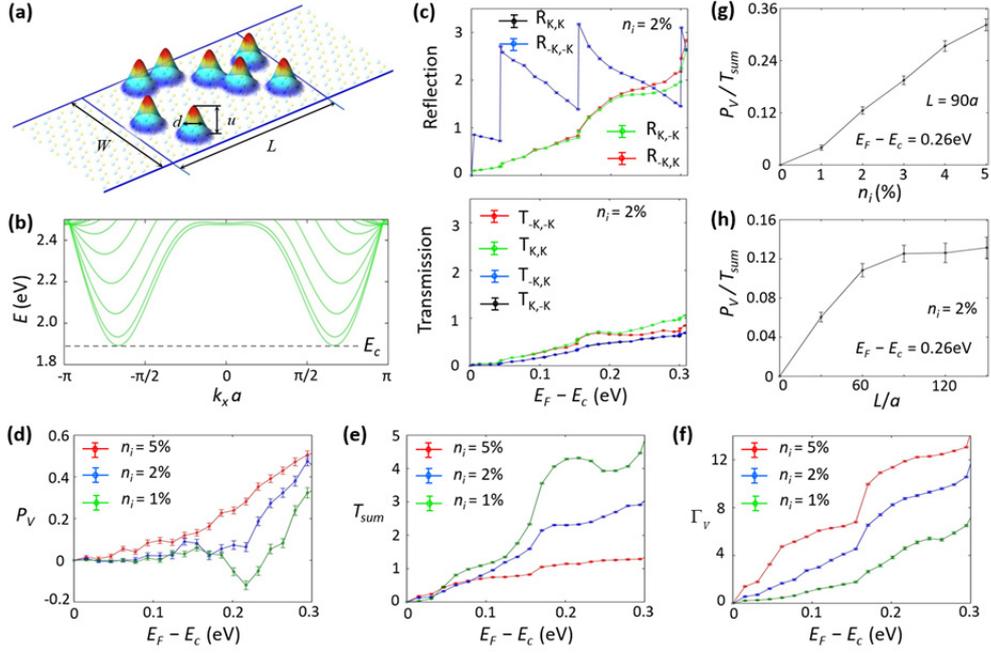

**Figure 2. Valley pump in monolayer MoS₂ nanoribbon.** (a) Gaussian disorders (see text) randomly distributed over the rectangular scattering region of length $L$ in a zigzag nanoribbon of width $W = 27.7a$ ($a$ is lattice constant). (b) Conduction subbands from the TB model without SOC. $E_c$ denotes the band edge. (c) The intra- and inter-valley transmission and reflection coefficients, as functions of the Fermi energy $E_F$. (d) The valley pump rate $P_V \equiv 2(R_{-K,K} - R_{K,-K})$, (e) the overall transmission $T_{sum} \equiv T_{K,K} + T_{-K,-K} + T_{K,-K} + T_{-K,K}$, and (f) the valley depolarization rate $\Gamma_V \equiv 2(T_{K,-K} + T_{-K,K} + R_{K,-K} + R_{-K,K})$. $L = 90a$ in (c-f), and $n_i$ is the disorder density. (g) The valley pump efficiency (i.e. per charge transmitted) $P_V/T_{sum}$ as a function of $L$, at $n_i = 2\%$. (h) $P_V/T_{sum}$ as a function of $n_i$, at $L = 90a$. $E_F - E_c = 0.26$eV in (g) and (h). In all plots, the dots represent calculated values averaged over 500 configurations of randomly generated disorder distributions, with the fluctuations shown as the error bar.

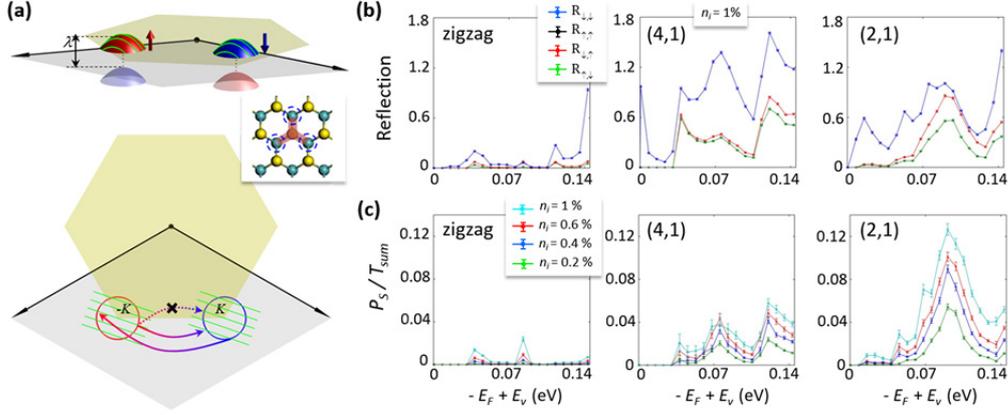

**Figure 3. Spin pump of spin-valley locked holes.** (a) Valence band edges in monolayer TMDs with the valley dependent spin splitting of magnitude $\lambda$. Red and blue denote spin up and down respectively. Green line cuts of the 2D bands give the 1D subbands of the nanoribbons (with periodic boundary condition). The two curved solid arrows denote the spin-flip reflections $R_{\uparrow,\downarrow}$ and $R_{\downarrow,\uparrow}$, which correspond to distinct momentum transfers on a finite Fermi surface with the spin-valley locking. Inset is the schematic of a non-magnetic disorder with spin-flip hopping between three nearest neighbor Mo sites. (b) Spin-conserved and spin-flip reflection coefficients calculated for monolayer MoS$_2$ nanoribbons of zigzag, (4,1) and (2,1) orientations respectively, at disorder density of $n_i = 1\%$. (c) The spin pump efficiency $P_S/T_{sum}$ at several disorder densities, where $P_S \equiv 2(R_{\uparrow,\downarrow} - R_{\downarrow,\uparrow})$. In (b) and (c), the widths of zigzag, (4,1) and (2,1) nanoribbons are $27.7a$, $27.5a$, and $29.1a$ respectively, and the length of the disordered regions are $90a$, $79.4a$, and $73.3a$ respectively. The dots represent average over 500 configurations of randomly generated disorder distributions, with the fluctuations shown as the error bar.